\documentclass[11pt]{article}
\pdfoutput=1
\usepackage{psfrag}
\usepackage{array}
\usepackage{amssymb}
\usepackage{amsmath}
\usepackage{amsthm}
\usepackage{float}
\usepackage{graphicx}
\usepackage{caption}
\usepackage{authblk}

\usepackage{epsfig}

\def\pb[#1,#2]{\{#1, #2\}}
\def\deb[#1,#2]{[#1,#2]_{\text{D.B.}}}

\def\Or[#1]{{\text{O}}\left({#1}\right)}
\def\dotl[#1,#2]{\left\langle #1,\, #2 \right\rangle}
\def\dotlb[#1,#2]{\left\langle #1,\, #2 \right\rangle}
\def\dotlm[#1,#2]{\left[ #1,\, #2 \right]}
\def\dotp[#1,#2]{(\vect{#1} \cdot\vect{#2})}
\def\aff[#1,#2]{\hat{#1}(#2)}

\def\n4sym{{\cal N}=4 SYM}
\def\>{\rangle}
\def\<{\langle}
\def\weight[#1,#2,#3]{\{(#1),#2,#3\}}
\def\ads[#1]{$\text{AdS}_{#1}$}

\newcommand{\be}{\begin{equation}}
\newcommand{\ee}{\end{equation}}
\newcommand{\ba}{\begin{align}}
\newcommand{\ea}{\end{align}}
\newcommand{\bs}{\begin{split}}
\def\sess\end{split}
\newcommand{\vect}[1]{{\boldsymbol{#1}}}

\title{Non-vacuum AdS cosmology and comments on gauge theory correlator }
\author{Soumyabrata Chatterjee\thanks{email: soumyac@iopb.res.in}} 
\author{Sudipto Paul Chowdhury\thanks{email: sudipto@iopb.res.in}}
\author{Sudipta Mukherji\thanks{email: mukherji@iopb.res.in}}
\affil{\normalsize Institute of Physics, \authorcr Sachivalaya Marg, \authorcr Bhubaneswar, India - 751 005. 
\authorcr and \authorcr Homi Bhabha National Institute, \authorcr Training School Complex, Anushakti Nagar, Mumbai, India
- 400 085}
\author{Yogesh K. Srivastava\thanks{email: yogeshs@niser.ac.in}}
\affil{\normalsize National Institute of Science Education and Research, \authorcr Jatni, Bhubaneswar, India - 752 050.
\authorcr and \authorcr Homi Bhabha National Institute, \authorcr Training School Complex, Anushakti Nagar, Mumbai, India
- 400 085}

\date{August, 2016}

\begin{document}
\maketitle

\begin{abstract}
 \noindent
Several time dependent backgrounds, with perfect fluid matter, can be used to construct solutions of Einstein 
equations in the presence of a negative cosmological constant along with some matter sources. In this
work we focus on the non-vacuum Kasner-AdS geometry and its solitonic generalization. To characterize these 
space-times, we provide ways to embed them in higher dimensional flat space-times. General space-like geodesics 
are then studied and used to compute the two point boundary correlators within the geodesic approximation.
\end{abstract}

\section{Introduction}

It has long been known that classical theory of gravity breaks down near space-time 
singularities and one would require an alternative description of gravity to explore such regions
in space-time. While such a 
complete description of gravity is still lacking, situation is considerably better when the gravitational 
theory has a description in terms of a non-gravitational dual. Indeed, in the past, several investigations were carried 
out to explore the physics of space-time singularities (behind the horizon or otherwise), where
the geometries permit such holographic gauge theory descriptions. For the black hole, where singularity is behind the 
horizon, many attempts have been made in past(see e.g., \cite{Kraus:2002iv}). For the present work, our focus will be on 
those situations which concern the cosmological singularities. 
These are the places where the time starts or ends and a natural question perhaps is to enquire if there is a way to resolve 
these singularities. While this is an important issue to address, it is also of importance to ask and identify the 
signatures of these cosmological
(space-like singularities) on their gauge theory duals. For some of the previous works on these issues, see 
\cite{Craps:2007ch} - \cite{Deger:2016axo}, and references therein.

One  of the simplest five dimensional backgrounds with a big bang 
singularity with a known gauge theory dual is the AdS-Kasner geometry. The singularity runs all 
the way to the boundary and  on this boundary we have a ${\cal N} = 4$ super Yang-Mills. Within this
set up, one may wish to analyze the nature of  boundary correlation functions -- particularly those
which may sense the singularity. Indeed, in \cite{Hertog:2005hu}, such a computation for two point correlation 
was performed for space-like operators of large dimensions in the geodesic approximation. 
It was found that if the direction along which the operators were separated carries a negative Kasner
exponent, the correlators developed divergences in their IR. Subsequently, in \cite{Banerjee:2015fua}, by solving
the space-like geodesic equations for arbitrary exponents, it was noticed that for correlators
separated along the positive Kasner exponents, correlation vanishes even if one pushes 
the space-like surface close to the singularity. While the reason for such behaviour is not
yet obvious to us, in the present work we consider a larger class of time dependent geometries
in AdS, study their symmetries, embeddings in flat space of higher dimensions, space-like geodesics in the geometries and  
nature two point correlators at the boundary.
Summary for the present work is described in the next paragraph.

In section $2$, we review and discuss various time dependent geometries in $4$ dimensions that will be of interest to us. Then in section $3$, 
we present flat space embeddings for flat AdS-FRW, AdS-Kasner and AdS-Kasner soliton geometries \footnote{AdS-Kasner soliton is an interesting
geometry in the sense that the horizon singularity associated with the AdS-Kasner is removed by adding an appropriate periodic coordinate.
This coordinate smoothly goes to zero in the bulk, capping off the geometry. AdS soliton was introduced in \cite{Horowitz:1998ha}
as the lowest energy bulk configuration for certain locally asymptotically AdS boundary conditions. Effect of the de-singularization was considered in
various contexts. See for example \cite{Haehl:2012fh}}.
In section $4$, we study $2$-point correlator 
in boundary theory for AdS-Kasner case using gauge-gravity duality. For some cases (in particular, AdS-Milne) case, we can explicitly solve 
the scalar field equations. Later we also calculate $2$-point correlator in geodesic approximation for AdS-Kasner and AdS-Kasner soliton 
cases using analytical and numerical techniques. Finally, we conclude with a discussion of future directions.

\bigskip

\section{Time dependent geometries in AdS}

\bigskip

In this section, we start with some well known time dependent geometries in four dimensional space-time (see for example, 
\cite{Erickson:2003zm}) both in the absence or in 
the presence of matter. Using previously known results, these  backgrounds can 
be used to construct a large class of time dependent
geometries which solve the Einstein equations in the presence of a negative cosmological constant and appropriate matter 
\cite{Fischetti:2014hxa}. Finally, in the next section, we embed all these time dependent geometries in higher dimensional flat space. Embeddings
of geometries in higher dimensional flat space illustrate many geometrical features of these spacetimes. 
Indeed, in \cite{Deser:1997ri}-\cite{Deser:1998xb},
thermal properties of genuinly static but curved geometries were understood from the thermal properties of accelerated observers  in corresponding
higer dimensional flat embeddings. Further, for time dependent Friedmann-Robertson-Walker geometries, embedding approach was
used to exibit three dimensional conformal symmetry on future space-like hypersurfaces \cite{Kehagias:2013xga}.
Though, the flat embeddings of our time 
dependent geometries are not  directly used in later sections, we believe such explicit constructions might be useful in 
future investigations and we include those in our subsequent discussions.

\bigskip

\noindent {\underbar{\bf Vacuum Solution}}

\bigskip

Vacuum Einstein equations have a simple anisotropic but homogeneous solution know as Kasner
solution. The Kasner metric is given by
\begin{equation}
dS^2 = - dt^2 + \sum_{i=1}^3 t^{2p_i} (dx^i)^2,
\label{kas}
\end{equation}
with
exponents $p_i$ satisfying the conditions
\begin{equation}
\sum_{i = 1}^3 p_i = 1,~~\sum_{i=1}^3 p_i^2 = 1.
\end{equation}
These $p_i$s can be parametrized by a single parameter $u \ge 1$ as
\begin{equation}
p_1 = -\frac{u}{1 + u + u^2}, ~p_2 = \frac{1+u}{1 + u + u^2}, ~{\rm and}~p_3 = \frac{u(1+u)}{1+u+ u^2}.
\label{vkas}
\end{equation}
These restrictions allow only two exponents to be positive and the other
negative except the case, where one of the $p_i$s is one and the rest are zero. This is known as a Milne
metric. The Milne metric covers only a part of the Minkowski space. This can be seen by making a 
simple coordinate transformation (say for $p_1 =1$, and $p_2 = p_3 = 0$)
\begin{equation}
y_0 = t~{\rm cosh} x_1, ~~ y_1 = t~{\rm sinh} x_1.
\label{makeflat}
\end{equation}
For other values of $p_i$s, Kasner metric has a curvature singularity at $t = 0$.

\bigskip

\noindent{\underbar{\bf In the presence of fluid matter}}

\bigskip

In the presence of fluid matter, there are several possibilities. In order to review, 
we start with the four dimensional Einstein equation 
\begin{equation}
\hat R_{\mu\nu} - \frac{1}{2} \gamma_{\mu\nu} \hat R = \hat T_{\mu\nu},
\label{eins}
\end{equation}
where the stress tensor is given by
\begin{equation}
\hat T_{\mu\nu} = (\rho + p) U_\mu U_\nu + p \gamma_{\mu\nu}.
\end{equation}
$U_{\mu}$ is the  four velocity of the fluid and $p, \rho$ are respectively the pressure and energy density of the 
fluid with the equation of state $p = w \rho$.
For such a system, the conservation of stress tensor  and the Einstein equations give
\begin{equation}
\rho = \frac{\rho_0}{t^2},
\end{equation}
(for some constant $\rho_0$. Weak energy condition would hold if $\rho_0 > 0$.) and
\begin{eqnarray}
&&p_1p_2 + p_1 p_3 + p_2p_3 = \rho_0,\nonumber\\
&&p_1 - p_1^2 + p_2 - p_1 p_2 - p_2^2 = \omega \rho_0,\nonumber\\
&&p_2 - p_2^2 + p_3 - p_2 p_3 - p_3^2 = \omega \rho_0,\nonumber\\
&&p_3 - p_3^2 + p_1 - p_1p_3 - p_1^2 = \omega \rho_0,\nonumber\\
\label{kk}
\end{eqnarray}
respectively.
To analyze the solutions, we introduce two parameters $S$ and $Q$ such that
\begin{eqnarray}
&&p_1 + p_2 + p_3 = S,\nonumber\\
&&p_1^2 + p_2^2 + p_3^2 = Q.
\label{kasner1}
\end{eqnarray}
In what follows, we review few possibilities that would be of our interest later.

When $S = Q = 1$ or equivalently, when $p = \rho = 0$, we have the anisotropic Kasner metric 
with $p_i$s given in (\ref{vkas}). With time, the metric expands in two directions and contracts along one.

For $\omega = 1$, pressure and energy density satisfy $p = \rho$. This leads to 
\begin{equation}
S = 1, ~~Q = 1 - 2\rho_0.
\end{equation}
Now, for a range of $\rho_0$, universe can have anisotropic non-contracting behaviour
in all the directions. To see this, we first express $p_2$ and $p_3$ in terms of $p_1$ are therefore,
\begin{eqnarray}
&&p_2 = \frac{1}{2} (1 - p_1 - {\sqrt{1 + 2 p_1 - 3 p_1^2 - 4 \rho_0}})\nonumber\\
&&p_3 = \frac{1}{2} (1 - p_1 + {\sqrt{1 + 2 p_1 - 3 p_1^2 - 4 \rho_0}}).
\end{eqnarray}
Now, for given $p_1 < 1$, for $p_2, p_3$ to be real
\begin{equation}
1 + 2 p_1 - 3 p_1^2 - 4 \rho_0 \ge 0.
\end{equation}
Or, in turn,
\begin{equation}
\rho_0 \le \frac{1}{4} (1 + 2p_1 - 3 p_1^2).
\end{equation}
For $0 < p_1 < 1$, the expression on the right varies from $1/2$ to $0$ picking at $p_1 = 1/3$
giving $\rho_0 = 1/3$. So, it is always possible to find a positive $p_1$ if
$\rho_0 \le 1/3$. Now, for $p_2$ to be positive, we need
\begin{equation}
1 - p_1 > {\sqrt{1 + 2 p_1 - 3 p_1^2 - 4 \rho_0}}.
\end{equation}
Or, in other words,
\begin{equation}
\rho_0 > p_1 (1 - p_1).
\end{equation}
The expression on the right varies between $0$ and $1/2$ peaking at $p_1 = 1/2$. 
Therefore, we see that if $\rho_0 < 1/3$, it is always possible to find $p_1$, such that 
$p_2$ is positive. Now, since $p_3 > p_2$, we see that, for $\rho_0$ below $1/3$, we will always
find all $p_i$s to be positive. Consequently, for the metric it means, all the directions
are anisotropically non-contracting starting from the initial big bang singularity. We note 
In particular,  a special case $p_1 = p_2 = 1/2, p_3 = 0$ with $\rho_0 = 1/4$
satisfy (\ref{kasner1}) which we will consider later. 

For other values of $\omega$, equations in (\ref{kk}) allow only FRW solutions.
We can then parametrize the isotropic metric with
\begin{equation}
p_1 = p_2 = p_3 = \alpha, ~S = 3\alpha, ~Q = 3 \alpha^2,
\label{lone}
\end{equation}
with
\begin{equation}
\alpha = \frac{2}{3 (1 + \omega)},~~\rho_0 = \frac{4}{3(1 + \omega)^2}.
\label{ltwo}
\end{equation}

\bigskip

\noindent{\underbar{\bf Time dependent AdS solutions}}

\bigskip

Now, following \cite{Fischetti:2014hxa}, all these four dimensional geometries can be understood as
solutions of the five dimensional gravity action in the presence of a negative cosmological constant and
supported by appropriate {\it anisotropic} matter stress tensor. Suppose we have metric $\gamma_{\mu\nu}$ and
perfect fluid matter with stress tensor $\hat T_{\mu\nu}$ satisfying four dimensional Einstein equation
(\ref{eins}). It is then possible to find a five dimensional metric $g_{ab}$ with matter stress
tensor $T_{ab}$ (with $a, b$ running from $0$ to $4$) which will satisfy five dimensional Einstein
equation
\begin{equation}
\label{enew}
R_{ab} - \frac{1}{2} g_{ab} R = -\Lambda g_{ab} + T_{ab},
\end{equation}
where $\Lambda$ is a negative cosmological constant. Here $g_{ab}$ and $T_{ab}$ are as follows:
\begin{eqnarray}
&&dS^2 = dy^2 + e^{2y/l} \gamma_{\mu\nu} dx^\mu dx^\nu,\nonumber\\
&&T_{\mu\nu} = \hat T_{\mu\nu}, \nonumber\\
&&T_{yy} = \frac{T}{2}, ~{\rm with}~T = S_{\mu \nu}T^{\mu \nu} = e^{-2y/l} \gamma_{\mu\nu} T^{\mu\nu}.
\end{eqnarray}
Here $l$ is related to the cosmological constant $\Lambda = - 6/l^2$. In the rest of the paper, we set $l=1$.
With a coordinate transformation $z = e^{-y}$, our five dimensional metric therefore becomes
\begin{equation}
dS^2 = \frac{1}{z^2}\Big( dz^2 - dt^2 +  \sum_{i=1}^3 t^{2p_i} dx_i^2\Big).
\label{metricc}
\end{equation}
While for unequal $p_i$s, we will call this as a non-vacuum AdS-Kasner metric, for equal $p_i$s we will call it an AdS-FRW  space. 

The five-dimensional energy-momentum tensor describes an anisotropic fluid. 
We now check whether this fluid satisfies the null$/$weak$/$strong energy condition(s). For an arbitrary null-vector $k_a$ in five-dimensions,
the null energy condition is given by $\mathcal{T}_{ab}k^a k^b \geq 0$, where 
$\mathcal{T}_{ab}= T_{ab}-\Lambda_{ab}$ is the total five-dimensional stress tensor (matter $+$ background). The null energy condition 
is satisfied by $\mathcal{T}_{ab}$ provided the parameter $\omega$ and 
the energy density $\rho$ appearing in the four-dimensional equation of state, satisfy the conditions $\omega \geq -\frac{1}{3}$ and $\rho \geq 0$ respectively. 
We recognise these as the conditions for the strong energy condition to be satisfied by the four-dimensional fluid. 
For the weak energy case, we first need to recognise that the weak energy condition may be violated by $\mathcal{T}_{ab}$
(Note that empty AdS also violates the weak energy condition). 
However, for the five-dimensional matter stress tensor $T_{ab}$ in (\ref{enew}), 
both the weak and the strong energy conditions \footnote{Recall that  the weak energy condition for the five-dimensional matter stress tensor is given by
\begin{equation}
\label{kasnerweak}
\widehat{T}_{\mu \nu} \xi^\mu \xi^\nu + T_{zz}(\xi^z)^2 \geq 0, 
\end{equation}
where $\xi^a$ is a five dimensional time-like vector and $u_\mu$ is the four-velocity of the boundary fluid. The strong energy condition for the same is given by 
\begin{equation}
\label{kasnertide}
\left(T_{a b}-\frac{T^{\prime}}{2} g_{a b}\right) v^a v^b \geq 0
\end{equation}
for an arbitrary future directed timelike vector field $v^a$, where $T^{\prime}= \frac{3}{2} T$}
are satisfied provided $\omega$ and $\rho$ satisfy the same conditions i.e $\omega \geq -\frac{1}{3}$ and $\rho \geq 0$.

As we pointed out earlier, we do not know if such a five dimensional set up  can be 
obtained naturally from ten or eleven dimensional supergravities. Consequently, a duality relation between the bulk 
gravity and the boundary gauge theory is not immediately obvious (except for $T_{ab} = 0$ where the bulk is a 
vacuum  AdS-Kasner metric). In the following, we will {\it assume} that such a duality relation exists and carry 
out our analysis.

The space-times discussed above generally suffer from singularities at the Poincare horizon $z = \infty$. This
singularity can be resolved by adding a compact direction to a Kasner-AdS. The metric is known as time-dependent
Kasner-AdS soliton and is given by
\begin{equation}
dS^2 = \frac{1}{z^2} \Big[ \Big(1 - \frac{z^5}{z_0^5}\Big)^{-1} dz^2 + \Big(1 - \frac{z^5}{z_0^5}\Big) d\theta^2
- dt^2 + \sum_{i=1}^3 t^{2p_i} dx_i^2\Big].
\label{ksoliton}
\end{equation}
The metric is nonsingular at $z = z_0$ if $\theta$ has a period $4 \pi z_0/5$.

\bigskip

\section{Flat space embeddings of AdS-cosmologies}

As mentioned in the introduction, it is often useful to embed curved metric into higher dimensional flat space. This illustrates various geometrical properties of spacetimes and  has been used in many previous works as cited in introduction.In this section,
we embed various cosmological metrics into flat space. 

\subsection{AdS-FRW embedding}
We first consider the AdS-FRW metric as it is simpler. We write it as
\begin{equation}
dS^2 = \frac{1}{z^2} \Big[dz^2 - dt^2 + a^2(t) \sum_{i=1}^3 dx_i dx^i\Big].
\label{newone}
\end{equation}
The dependence of the scale factor $a(t)$ on $t$ follows from Einstein equations and generally goes as $t^p$ 
for some constant $p$. Let us now consider all the terms inside the square brackets except the $dz^2$ part.
Introducing new coordinates \cite{Paston:2013uia}, 
\begin{eqnarray}
&&y^i = a(t) x^i,~~i = 1, 2, 3, \nonumber\\
&&y^0 = c a(t) \sum_{i=1}^3 x^i x^i + c \int \frac{dt}{\dot a(t)} + \frac{a(t)}{4c},\nonumber\\
&&y^4 = c a(t) \sum_{i=1}^3 x^i x^i + c \int \frac{dt}{\dot a(t)} - \frac{a(t)}{4c}.
\end{eqnarray}
with $c$ being an arbitrary constant, we get
\begin{equation}
- dt^2 + a^2(t) \sum_{i=1}^3 (dx^i)^2 = - (dy^0)^2 + \sum_{i=1}^4 (dy^i)^2.
\end{equation}
Consequently, the metric (\ref{newone}) reduces to AdS in Poincare coordinates,
\begin{equation}
dS^2 = \frac{1}{z^2} \Big[ dz^2 - (dy^0)^2  + \sum_{i=1}^4 (dy^i)^2\Big].
\end{equation}
This can now easily be embedded in flat space using the standard map
\begin{eqnarray}
&&X_0 = \frac{1}{2z} \Big(z^2 + 1 + \sum_{i=1}^4 (y^i)^2 - (y^0)^2\Big),\nonumber\\
&&X_i = \frac{y^i}{z}, ~~~ i = 1, 2, 3, 4,\nonumber\\
&&X_5 =  \frac{1}{2z} \Big(z^2 - 1 + \sum_{i=1}^4 (y^i)^2 - (y^0)^2\Big),\nonumber\\
&&X_6 = \frac{y^0}{z}.
\end{eqnarray}

To summarize, the five dimensional AdS-FRW metric can be embedded in seven dimensional flat space
\begin{equation}
dS^2 = - dX_0^2 - dX_6^2 + \sum_{i=1}^{5} dX_i^2,
\end{equation}
with
\begin{eqnarray}
&&X_0 = \frac{1}{2 z} \Big[ z^2 + 1 - a(t) \int \frac{dt}{\dot a(t)}\Big],\nonumber\\
&&X_j = \frac{a(t) x^j}{z}, ~~j = 1, 2, 3,\nonumber\\
&& X_4 = \frac{1}{z} \Big[ c a(t) \sum_{j=1}^3 x_j^2 + c \int \frac{dt}{\dot a(t)} - \frac{a(t)}{4 c}\Big], \nonumber\\
&& X_5 = \frac{1}{2z} \Big[ z^2 - 1 - a(t) \int \frac{dt}{\dot a(t)}\Big],\nonumber\\
&&X_6 = \frac{1}{z} \Big[c a(t) \sum_{j=1}^3 x_j^2 + c \int \frac{dt}{\dot a(t)} + \frac{a(t)}{4 c}\Big].
\end{eqnarray}
>From the higher dimensional perspective, the AdS-FRW metric sits on the intersection of the hypersurfaces
\begin{equation}
-X_0^2 - X_6^2 + \sum_{1}^5 X_i^2 = -1,
\end{equation}
and
\begin{equation}
\frac{1- X_{0}^{2} + X_{5}^{2}}{(X_0 -X_5)^2} = 
(a(t)\int \frac{dt}{\dot a(t)} )\Big|_{\substack{a(t)=2c\frac{X_6 -X_4}{X_0 -X_5}}}.
\end{equation}

Particularizing to the case $a(t)= t^p$, we get 
\begin{equation}
2 c (X_6 - X_4) (X_0 - X_5)^{(p-1)} = \Big( (2 - p) p (1 - X_0^2 + X_5^2)\Big)^{\frac{p}{2}}.
\end{equation}

\subsection{Embedding AdS-Kasner}
For the flat embedding of AdS-Kasner, it requires a bit more work. Consider the metric in 
(\ref{metricc}). We first take the part
\begin{equation}
-dt^2 + t^{2p_1} (dx^1)^2.
\end{equation}
This is in the FRW form. Therefore, using our previous results, we can embed this part to flat
space. We introduce 
\begin{eqnarray}
&&y^0 = (c (x^1)^2 + \frac{1}{4c}) t^{p_1} + \frac{c t^{2-p_1}}{p_1 (2 - p_1)},\nonumber\\
&&y^1  = t^{p_1} x^1,\nonumber\\
&&\tilde y =  (c (x^1)^2 - \frac{1}{4c}) t^{p_1} + \frac{c t^{2-p_1}}{p_1 (2 - p_1)}.
\end{eqnarray}
Then (\ref{metricc}) takes the form
\begin{equation}
dS^2 = \frac{1}{z^2}\Big[ dz^2 - (dy^0)^2 + (dy^1)^2 + (d\tilde y)^2 + t^{2p_2} dx^2 dx^2 + t^{2p_3} dx^3 
dx^3\Big].
\label {metriccc}
\end{equation}
Let us now define light-cone coordinates $u = y^0 - \tilde y, v = y^0 + \tilde y$ to write (\ref{metriccc})
as
\begin{equation}
dS^2 = \frac{1}{z^2} \Big[dz^2 - du dv + (dy^1)^2 + (2 c u)^{\frac{2p_2}{p_1}} (dx^2)^2+ (2 c 
u)^{\frac{2p_3}{p_1}} (dx^3)^2 \Big].
\end{equation}

So Kasner can be embedded as a hypersurface  
\begin{equation}
uv -\tilde{y}^2 = \frac{1}{p_1 (2-p_1)}(2cu)^{\frac{2}{p_1}},
\end{equation}
in the plane wave metric. It is useful now to convert the above plane wave metric into Brinkmann form by defining
\begin{eqnarray}
&&y^2 = (2cu)^{\frac{p_2}{p_1}} x^2,~~y^3 = (2cu)^{\frac{p_3}{p_1}} x^3, ~~U = u,\nonumber\\
&&V = v + \frac{2cp_2}{p_1} (2 c u)^{\frac{2p_2 -p_1}{p_1}} (x^2)^2 +  \frac{2cp_3}{p_1} (2 c u)^{\frac{2p_3 
-p_1}{p_1}} (x^3)^2.
\end{eqnarray}
The metric becomes
\begin{equation}
dS^2 = \frac{1}{z^2} \Big[ dz^2 - dU dV + \sum_{i=1}^3 dy^i dy^i - H dU^2\Big],
\end{equation}
where
\begin{eqnarray}
H &&= \frac{ (2cU)^{\frac{2p_2}{p_1}} p_2 (p_1 - p_2) (x^2)^2 +  (2cU)^{\frac{2p_3}{p_1}} p_3 (p_1 - p_3) 
(x^3)^2 }{p_1^2 U^2} \noindent\\
&&=\frac{ ( p_2 (p_1 - p_2) (y^2)^2 +  p_3 (p_1 - p_3) 
(y^3)^2 }{p_1^2 U^2}.
\end{eqnarray}

In the Brinkmann form, hypersurface equation can be written as
\begin{equation}
UV -\frac{p_2}{p_1}(y^2)^2 -\frac{p_3}{p_1}(y^3)^2  -(y^1)^2 = \frac{1}{p_1 (2-p_1)}(2cU)^{\frac{2}{p_1}}.
\end{equation}
Further, introducing 
\begin{eqnarray}
&&Z_0 = \frac{V + U + HU}{2},~Z_4 = \frac{V-U + HU}{2}, ~Z_5 = \frac{H + \frac{U^2}{2}}{2},\nonumber\\
&&Z_6 = \frac{H - \frac{U^2}{2}}{2}, ~Z_i = y^i ~(i =1, 2, 3),
\end{eqnarray}
we get flat space embedding of plane wave. So $4$-dimensional Kasner can be embedded in $7$ dimensional flat space. Using this, we get 
\begin{equation}
dS^2 = \frac{1}{z^2} \Big[ dz^2 - dZ_0^2 - dZ_6^2 + \sum_{i=1}^5 dZ_i^2\Big].
\label{flatads}
\end{equation}
This is now AdS in Poincare form and can be easily embedded in flat space.
So to summarize, the five dimensional AdS-Kasner metric can be embedded in flat space 
\begin{equation}
dS^2 = - dX_0^2 - dX_7^2 - dX_8^2 + \sum_{i=1}^{6} dX_i^2,
\end{equation}
and to write it in a compact way, we first define
\begin{eqnarray}
&&Z_0 = \frac{t^{p_1}}{4 c} + c \Big[ \frac{t^{2-p_1}}{(2 - p_1)p_1} 
+ \frac{1}{p_1^2} ( p_1^2 t^{p_1} x_1^2 + (2p_1 - p_2) p_2 t^{2 p_2 -p_1} x_2^2 + 
(2p_1 - p_3)p_3 t^{2p_3 - p_1}x_3^2)\Big],\nonumber\\
&&Z_1 = x_1 t^{p_1}, ~~Z_2 = x_2 t^{p_2}, ~~Z_3 = x_3 t^{p_3},\nonumber\\
&&Z_4 = -\frac{t^{p_1}}{4 c} + c \Big[ \frac{t^{2-p_1}}{(2 - p_1)p_1}        
+ \frac{1}{p_1^2} ( p_1^2 t^{p_1} x_1^2 + (2p_1 - p_2) p_2 t^{2 p_2 -p_1} x_2^2 +
(2p_1 - p_3)p_3 t^{2p_3 - p_1}x_3^2)\Big],\nonumber\\
&&Z_5 = \frac{t^{2p_1}}{16 c^2} + \frac{2c^2}{p_1^2}
\Big[(p_1 - p_2) p_2 t^{2(p_2 - p_1)} x_2^2 + (p_1 - p_3)p_3 t^{2(p_3 - p_1)} x_3^2\Big]\nonumber\\
&&Z_6 = -\frac{t^{2p_1}}{16 c^2} + \frac{2c^2}{p_1^2}
\Big[(p_1 - p_2) p_2 t^{2(p_2 - p_1)} x_2^2 + (p_1 - p_3)p_3 t^{2(p_3 - p_1)} x_3^2\Big].
\end{eqnarray}
The flat coordinates are then
\begin{eqnarray}
&&X_0 = \frac{1}{2z}\Big[z^2 + 1 + \sum_{i=1}^5 Z_i^2 - Z_0^2 - Z_6^2\Big],\nonumber\\
&&X_i = \frac{Z_i}{z},~~i = 1, ..., 5, \nonumber\\
&&X_6 = \frac{1}{2z}\Big[z^2 - 1 + \sum_{i=1}^5 Z_i^2 - Z_0^2 - Z_6^2\Big],\nonumber\\
&&X_7 = \frac{Z_0}{z},\nonumber\\
&&X_8 = \frac{Z_6}{z}.
\end{eqnarray}
Here $c$ is an arbitrary constant. 

AdS Kasner is then an intersection of following hypersurfaces in flat space.
\begin{eqnarray}
&&-X_0^{2} -X_7^{2} - X_8^2 + X_6^2 + \sum_{i=1}^{5} X_i^2 = -1, \nonumber\\
&&(X_7 -X_4)^2 = 2(X_5 -X_8)(X_0 -X_6), \nonumber\\
&&X_5^2 -X_8^2 = \frac{p_2(p_1 -p_2)}{2p_1^2} X_2^2 + \frac{p_3(p_1 -p_3)}{2p_1^2} X_3^2,  \nonumber\\
&&X_7^2 -X_4^2 -2X_5^2 + 2X_8^2 -\frac{p_2}{p_1} X_2^2 - \frac{p_3}{p_1} X_3^2 \nonumber \\ 
&&= (X_0 -X_6)^2\left(\frac{1}{(2-p_1)p_1}\left(2c \frac{X_7-X_4}{X_0 -X_6}\right)^{\frac{2}{p_1}}+ \left(\frac{X_1}{X_0-X_6}\right)^2\right).
\end{eqnarray}
\subsection{AdS-Kasner soliton}
Finally, we embed the Kasner-AdS soliton (\ref{ksoliton}) in flat space-time. First, we embed the 
Kasner part of the metric as was done in equation (\ref{flatads}). We get
\begin{equation}
dS^2 = \frac{1}{z^2} \Big[ \Big(1 - \frac{z^5}{z_0^5}\Big)^{-1} dz^2 + \Big(1 - \frac{z^5}{z_0^5}\Big) d\theta^2
 - dZ_0^2 - dZ_6^2 + \sum_{i=1}^5 dZ_i^2\Big].
\end{equation}
Now, with the following choice of coordinates
\begin{eqnarray}
&&X_0 = \frac{Z_0}{z},~~X_1 = \frac{Z_6}{z},~~X_2=\frac{2 z_0}{5}\sqrt{g(z)}~{\rm Sin}\left(\frac{5\theta}{2 z_0}\right),~~
X_3=\frac{2 z_0}{5}\sqrt{g(z)}~{\rm Cos}\left(\frac{5\theta}{2 z_0}\right), \nonumber\\
&&X_4=\frac{Z_1}{z},~~X_5=\frac{Z_2}{z},~~X_6=\frac{Z_3}{z},~~X_7=\frac{Z_4}{z},~~X_8=\frac{Z_5}{z}\nonumber\\
&&X_9 = \frac{1}{2z}\left[ Z_0^2 + Z_6^2 - \sum_{i=1}^5 Z_i^2 + z D(z) +1\right],\nonumber\\
&&X_{10} =  \frac{1}{2z}\left[ Z_0^2 + Z_6^2 - \sum_{i=1}^5 Z_i^2 + z D(z) -1\right],
\end{eqnarray}
the metric reduces to
\begin{equation}
dS^2 = -dX_0^2-dX_{10}^2-dX_1^2+\sum_{i=2}^9 dX_i^2
\end{equation}
In the above
\begin{equation}
g(z)=\frac{1}{z^2}\left(1-\frac{z^5}{z_0^5}\right), ~~ D(z)=-\int\frac{dz}{z^2g(z)}\left(1-\frac{z_0^2z^4 g'(z)^2}{25}\right).
\end{equation}
The soliton metric therefore sits at the intersection of six hypersurfaces. It is straightforward to find the surfaces for 
the soliton. Computations are similar to that of AdS-Kasner and we do not reproduce it here.

\section{On the Gauge theory correlators}
In this section our aim is to study the two point correlators of the operators in the boundary gauge theory
for time-dependent geometries. 

\subsection{Using Bulk Dual}
One conventional way to compute the two-point functions is to use
the bulk dual. We solve the bulk equations of motion of a  massive scalar whose boundary value acts as a
source for the operators of the boundary theory. Varying the boundary action (evaluated on the solution)
twice with respect to the source, we get the two point functions of the operators.
In general, in time dependent backgrounds, the energy of the scalar field does not remain constant. This
introduces problems in choosing boundary conditions and subsequently defining a vacuum. However, in some cases progress can be 
made. In the first part of this section, we review the situation.
 
The equation of motion of a massive scalar field is given by
\begin{equation}
(\square- m^2) \phi =0.
\label{cov}
\end{equation}
where $\Box = \frac{1}{\sqrt{-g}}\partial_a({\sqrt{-g}}g^{ab} \partial_b \phi)$.
Using the fact that for the metric (\ref{metricc}),
\begin{equation}
{\sqrt{-g}} = \frac{t^{p_1+p_2+p_3}}{z^5} = \frac{t}{z^5},
\end{equation}
we get
\begin{equation}
z^2 \partial_z^2 \phi - 3 z \partial_z \phi - z^2 \partial_t^2\phi - \frac{z^2}{t} \partial_t\phi 
+ \sum_{i=1}^3 \frac{z^2}{t^{2p_i}} {\partial_i}^2\phi - m^2 \phi = 0.
\end{equation}
Substituting the ansatz,
\begin{equation}
\phi = \sum_{i=1}^3\xi(t) \rho(z)e^{ik_i x^i},
\end{equation}
we get
\begin{eqnarray}
&&z^2\partial_z^2\rho - 3 z \partial_z \rho - (m^2 + \omega^2 z^2)\rho = 0, \nonumber\\
&&t\partial_t^2\xi + \partial_t\xi +  \sum_{i=1}^3 \frac{k_i^2}{t^{2p_i -1}}\xi - \omega^2 t \xi =0.
\label{eqn}
\end{eqnarray} 
Here $\omega$ is an arbitrary constant. Now, first is the usual radial equation in vacuum AdS. 
The solutions can be written in terms of Bessel functions.
In order to solve the time dependent part, it is instructive to introduce a new variable $\chi(t)$ as
\begin{equation}
\xi = t^{-\frac{1}{2}} \chi(t),
\end{equation}
and re-write the second equation in (\ref{eqn}) as
\begin{equation}
\partial_t^2 \chi + \Big(\frac{1}{4t^2} + \sum_{i=1}^3 \frac{k_i^2}{t^{2p_i}} - \omega^2\Big) \chi = 0.
\label{eqnone}
\end{equation}
Normalization can be fixed by setting the the Wronskian to a constant, for example, by choosing
\begin{equation}
\chi \partial_t \chi^* - \chi^* \partial_t \chi = i.
\label{normal}
\end{equation}
If we treat (\ref{eqnone}) as a Schrodinger-like equation, the effective potential has a form
\begin{equation}
V(t) = - \frac{1}{4t^2} - \sum_i \frac{k_i^2}{t^{2p_i}}.
\end{equation}  
Near $t = 0$, for generic $p_i$s, the first term dominates making the potential singular and scale 
invariant.

The case with one of the $p_i$s equals  one and the rest zero, say $p_1 =1, p_2 = p_3 = 0$, the space-time 
is non-singular. With the coordinate transformations given in (\ref{makeflat}),
it can be make flat. Hence the scalar field equations and, consequently, finding the boundary correlator
becomes simple. However, solving it directly requires some work and, since we have not seen any explicit calculations
in the literature, we briefly present it here. It is useful to first analytically continue the coordinates to 
$t = i \tau$ and $x_1 = i \psi$. The geometry has a horizon at $\tau = 0$ and the boundary at $z = 0$.
If we insist the field to vanish at these two points\footnote{We will see later that, with this boundary conditions, the 
final result and the one calculated from the geodesic approximation matches.}, the solution can be written as
\begin{equation}
\phi = \sum_{n = -\infty}^{\infty} \epsilon^{\Delta_-} \int _0^{\infty} \int_0^{\infty} d\omega dk_2 dk_3
C_{\omega, n, k_2, k_3} \sqrt{\tilde \omega} e^{in\psi} e^{i k_2 x_2} e^{ik_3 x_3}
\frac{z^2 K_\alpha (\tilde \omega z)}{\epsilon^2 K_\alpha(\tilde \omega \epsilon)} J_n (\omega \tau).
\end{equation}
Here, $J_n$ is the Bessel function, $K_\alpha$ is a modified Bessel function, $\epsilon$ is a small cutoff near $z = 0$ and 
$C_{\omega, n, k_1, k_2}$ are  constants. The other constants are defined as 
$\tilde \omega = {\sqrt{\omega^2 + k_1^2 + k_2^2}}$, $\alpha = \sqrt{4 + m^2}$ and $\Delta_- = 2 - \alpha$. 
Now, following the standard prescription, we evaluate the scalar field
action on the solution. This gives,
\begin{eqnarray}
&&S = \sum_{n_1, n_2} \int d\tau dx_2 dx_3 d\omega d\omega^\prime dk_2 dk_3 dk_2^\prime dk_3^\prime \delta_{m + n}
e^{i(k_2 + k_2^\prime)x_2} e^{i (k_3 + k_3^\prime) x_3}\nonumber\\
&&~~~~~~~~~~~~C_{\omega, n, k_2, k_3}C_{\omega^\prime, m, k_2^\prime, k_3^\prime} 
\tau J_{n} (\tau \omega) J_{m} (\tau \omega^\prime) {\sqrt{\omega \omega^\prime}} {\tilde \omega}^{2\alpha}.
\end{eqnarray}
Further, differentiating the above action twice with respect to $C$, we reach at the following two point correlator
in momentum space:
\begin{equation}
\langle{\cal O}(k_i, \omega, n) {\cal O} (k_i^\prime, \omega^\prime, m)\rangle
\sim (-)^m \delta_{n+m} \delta(k_2 + k_2^\prime) \delta (k_3 + k_3^\prime) \delta (\omega- \omega^\prime)
\tilde \omega^{2\alpha}.
\end{equation}
Here, we have used
\begin{equation}
\int_0^\infty d\tau \tau J_n(\omega \tau) J_n (\omega^\prime \tau) = \frac{1}{\omega} \delta(\omega - \omega^\prime).
\end{equation}
To obtain the correlator in position space, we use the identity
\begin{equation}
\sum_{n=-\infty}^{\infty} J_n(\omega \tau) J_n(\omega \tau^\prime) e^{in (\psi - \psi^\prime)} = J_0 (\omega \tilde\tau)
\end{equation}
where
\begin{equation}
\tilde\tau = {\sqrt{\tau^2 + \tau^{\prime 2} - 2 \tau \tau^\prime~{\rm cos}(\psi - \psi^\prime)}},
\end{equation}
and get
\begin{equation}
\langle{\cal O}(\tau, \psi, x_2, x_3) {\cal O} (\tau^\prime, \psi^\prime, x_2^\prime, x_3^\prime)\rangle
= \int_{0}^\infty d\omega \int_{-\infty}^\infty dk_2 \int_{-\infty}^\infty dk_3
~\omega {\tilde \omega}^{2 \alpha} J_0 (\omega \tilde \tau)e^{i k_2 (x_2 - x_2^\prime)} e^{i k_3 (x_3 - x_3^\prime)}.
\end{equation}
Going to the polar coordinates the integrals can be easily performed and we arrive at
\begin{equation}
\langle{\cal O}(\tau, \psi, x_2, x_3) {\cal O} (\tau^\prime, \psi^\prime, x_2^\prime, x_3^\prime)\rangle
= \frac{1}{[\tau^2 + \tau^{\prime 2} - 2 \tau \tau^\prime ~{\rm{cos}}(\psi - \psi^\prime)
+ (x_2 - x_2^\prime)^2 + (x_3 - x_3^\prime)^2]^{2 + \alpha}},
\end{equation}
up to some $\alpha$ dependent prefactor.
Now, analytically continuing back to the earlier coordinates, we reach at
\begin{equation}
\langle{\cal O}(t, x_1, x_2, x_3) {\cal O} (t^\prime, x_1^\prime, x_2^\prime, x_3^\prime)\rangle
= \frac{1}{[- t^2 - t^{\prime 2} + 2 t t^\prime ~{\rm{cosh}}(x_1 - x_1^\prime)
+ (x_2 - x_2^\prime)^2 + (x_3 - x_3^\prime)^2]^{2 + \alpha}}.
\label{onetwo}
\end{equation}
For space-like separated points, say by a distance $2x$ along $x_1$, the formula
simplifies to
\begin{equation}
\langle{\cal O}(t, x, x_2, x_3) {\cal O} (t, -x , x_2, x_3)\rangle
= \frac{1}{(2 t ~{\rm {sinh}} x)^{2 \Delta}},
\label{twotwo}
\end{equation}
where $\Delta = 2 + \alpha$.

For generic values of $p_i$s, the $\xi(t)$ in (\ref{eqn}) can not be solved analytically. However, for specific 
cases, some progress can be made. To illustrate this further, let us consider the case $\{p_1 = \frac{1}{2},
p_2 = \frac{1}{2}, p_3 =0\}$. As discussed in the previous section, for $\rho_0 = \frac{1}{4}$, the metric
solves the Einstein equation. Here one gets,
\begin{eqnarray}
\chi(t) &&= C_1 M[-i \beta, 0, 2 i \alpha t] + C_2 W[-i \beta, 0, 2 i \alpha t],~ {\rm for}~k_3^2 >
\omega^2\nonumber\\
&&= C_3 M [\beta, 0, 2 \alpha t] + C_4 W[\beta, 0, 2 \alpha t], ~ {\rm for}~k_3^2 < \omega^2.
\label{soln}
\end{eqnarray}
Here
\begin{equation}
\alpha = {\sqrt{|k_3^2 - \omega^2|}}, ~ \beta = \frac{k_1^2 + k_2^2}{2\alpha}.
\end{equation}
$M$, $W$ are the Whittaker functions and $C_i$s are integration constants. Owing to the properties of
$M$ and $W$, we note that, at $t = 0$, the solutions identically vanish. Further, because of time-dependence
of the metric, there is no natural way to identify the positive and negative energy modes. Therefore we find
it difficult to put appropriate boundary conditions to isolate the relevant one form the general solution.

\subsection{Geodesic approximation}

Another way to approximate the  two point correlator of {\it higher} dimensional operators is
to relate it with geodesic computations. Within the semi-classical limit, this approximation suggests that the
boundary correlator of two operators can be written as
\begin{equation}
\langle \psi|
{\cal O}(x) {\cal O}(x^\prime) |\psi \rangle
\sim e^{-m {\cal L}_{reg}(x, x^\prime)},
\label{cor}
\end{equation}
where $|\psi\rangle$ is some state of the boundary field theory and ${\cal O}$ is an
operator whose dimension $\Delta$ is large and close to $m$, the
mass of a heavy bulk
field.  For our purpose ${\cal O}$ will be a scalar operator and, consequently, we will
consider only the scalar field. The quantity
${\cal L}_{reg}$ is the length of the regularized bulk geodesic connecting points $x$ and $x^\prime$
\footnote{Here one represents the inverse of the operator $\Box - m^2$ as a path integral. In
the large mass limit and in the saddle point approximation, the path integral is then expected to reduce
to finding the geodesic. This method works well for Euclidean time. However, for Lorentzian metric (as in
the case here), there are often subtleties due to the appearance of oscillating phases. In our discussion, we
will {\it assume} such uncertainties \cite{Louko:2000tp} - \cite{Festuccia:2005pi} do not occur.}.

Let us, for illustrative purpose, start with the Milne universe. In order to find the 
space-like two point correlator, we need to find the geodesic whose end points are stuck at the boundary, say at
$t = t_0$. We take the points along $x_1$ direction at $x_1 = \pm x$.
The geodesic equations are
\begin{eqnarray}
&&\ddot x_1 + 2 \frac{\dot x_1}{t} - t \dot x_1^3 = 0,\nonumber\\
&& z \ddot z + \dot z^2 + t^2 \dot x_1^2 - t z \dot z \dot x_1^2 - 1 = 0.
\end{eqnarray}
Here, the dots are derivatives with respect to $t$.
The solutions of the above equations are 
\begin{equation}
x_1 = \pm ~{\rm{log}}~\Big[\frac{{\sqrt{c}} t}{1 + {\sqrt{1 - c t^2}}}\Big],~~~z = {\sqrt{t^2 - t_0^2}}.
\end{equation}
The constant $c$ is related to the turning point of the geodesic in the bulk. Since at this point,
$x_1 = 0, dt/dx = 0$, and therefore $t = 1/{\sqrt c}$. Further, since at $ t = t_0$, $x_1 = \pm x$,
we get 
\begin{equation}
c = \frac{4 e^{2x}}{(1 + e^{2x})^2 t_0^2}.
\label{cond}
\end{equation}
For calculational simplicity, without loss of generality, we take $t_0 =1$ henceforth. 
The length of the geodesic can be easily computed. It is
\begin{equation}
{\cal {L}} = \int_{t=\frac{1}{\sqrt c}}^{1 + t(z = \epsilon)}
\frac{2 dt}{z} {\sqrt{-1 + t^2 \dot x^2 + \dot z^2}}.
\end{equation}
The lower integration limit is the turning point of the geodesic and the upper limit
is the value of $t$ at which one approaches the boundary $z = \epsilon$ with $\epsilon \rightarrow 0$.
The result diverges as we take $\epsilon$ to zero. The regularization is done by subtracting the equivalent 
piece coming from AdS. This gives
\begin{equation}
{\cal L}_{reg} = ~{\lim}_{\epsilon \rightarrow 0}~\left(\rm{log}~\Big[\frac{4 ~{\rm {sinh}}^2 x}{\epsilon}\Big] - ~2~\rm{log}\Big[\frac{1}{\epsilon}\Big]\right)
= ~{\rm {log}}[4 ~{\rm {sinh}}^2 x].
\end{equation}
Consequently,
\begin{equation}
\langle {\cal O} (1, x, 0, 0) {\cal O} (1, -x, 0, 0)\rangle = \frac{1}{(2~{\rm {sinh}}x)^{2m}}.
\end{equation}
This is same as (\ref{twotwo}) for the operators of large dimensions.  

\bigskip

\noindent {\underbar{\bf General spacelike geodesics}}

\bigskip

We now focus on to space-like geodesics in more general Kasner-AdS space-time. 
Our computation will equally go through for non-vacuum solution. 
Here, specifically, we look for the 
geodesics for which the end points are situated at $(x_1, x_2, x_3)$ and $(- x_1, - x_2, 
- x_3)$ at time $t = t_0$. Taking $t$ as a parameter, the geodesic equations are\footnote{In earlier papers, 
analysis were done for the correlators along one particular direction. Computational difficulty reduces 
significantly in those cases.}
 
\begin{eqnarray}
&&\ddot x^i + 2 p_i \frac{\dot x^i}{t} - \dot x^i \sum_{j=1}^3 p_i t^{2p_j -1} (\dot x^j)^2 = 0. \nonumber\\
&&z \ddot z + \dot z^2 + \sum_{j=1}^3 t^{2p_j}(\dot x^j)^2 - z \dot z \sum_{j = 1}^3 p_jt^{2p_j -1} (\dot x^j)^2 - 1 = 0.
\label{compli}
\end{eqnarray}

Though we have complicated coupled set of equations, it turns out that much can be said about the solutions without assuming
explicit values of $p_i$s. We first define 
\begin{equation}
y^i = \dot x^i, ~~~K = z \dot z,
\end{equation}
and rewrite (\ref{compli}) as 
\begin{eqnarray}
&&\dot y^i + \frac{2p_i}{t} y^i - y^i  \sum_{j=1}^3 p_i t^{2p_j -1} (y^j)^2 = 0. \nonumber\\
&&\dot K - K  \sum_{j = 1}^3 p_jt^{2p_j -1} (y^j)^2 +  \sum_{j=1}^3 t^{2p_j}(y^j)^2 - 1 = 0.
\label{complione}
\end{eqnarray}
The first set of equations integrate to
\begin{equation}
y^i = C_{ij} t^{-2(p_i - p_j)} y^j,
\label{complifour}
\end{equation}
where $C_{ij}$s are the integration constants satisfying $C_{ji} = C_{ij}^{-1}$.
We now assume it there is a turning point of the geodesic and we take that to
be $t = t^*$. Using (\ref{complifour}) in the first equation of (\ref{complione}) and solving for $y^i$, we get
\begin{equation}
y^i = \pm \frac{{t^{-2p_i}}}{  {\sqrt{  t^{-2p_i} - 3{t^*}^{-2p_i} + \sum_{j \ne i} {t^*}^{-2(p_i-p_j)} t^{-2p_j}}}    }.
\end{equation}
To fix the integration constants, we have used the boundary condition $dt/dx^i = 0$ at $t = t^*$ and
also the fact that the equations in (\ref{complione}) have certain scaling symmetry. To elaborate further,
we note that under the transformation 
\begin{equation}
z = \lambda \bar z, ~~t = \lambda \bar t, ~~x^i = \lambda^{1-p_i} \bar x^i,
\label{scal}
\end{equation}
with $\lambda$ constant, 
the forms of the equations do not change. This alone can be used to fix all the $C_{ij}$s.
We get
\begin{equation}
C_{ij} = {t^*}^{p_i - p_j}.
\end{equation}
Hence,
\begin{equation}
x^i = \pm \int  dt~~\frac{{t^{-2p_i}}}{  {\sqrt{  t^{-2p_i} - 3{t^*}^{-2p_i} + \sum_{j \ne i} {t^*}^{-2(p_i-p_j)} t^{-2p_j}}}    }.
\label{complitwo}
\end{equation}
The constants arising from these integrals are to be fixed using the boundary conditions $x^i = 0$ at $t = t^*$.
We can now use this to rewrite $K$ (and therefore $z$) of the second equation in (\ref{complione}) 
in the form
\begin{equation}
z(t)=  {\sqrt{ \int \frac{2}{g_1(t)}\Big[ \int_1^t d \tau~~ \Big(1 - g_2(\tau)\Big) g_1(\tau) + C_1\Big] dt + C_2   }},
\label{complifive}
\end{equation}
with
\begin{equation}
g_1(t) = e^{- \int^t (\sum_{j=1}^3p_j \tau^{2p_j -1} {y^j}^2) d\tau}, ~~g_2(t) = \sum_{j=1}^3 t^{2p_j} {y^j}^2.
\end{equation}
In (\ref{complifive}), $C_{1,2}$ are the integration constants which are to be fixed using $z =0$ at $t = t_0$ and 
$z$ has a maximum at $t = t^*$.  

To proceed further, one needs to use explicit values of $p_i$s. For some set of values, the integrations can be analytically
performed and for the rest, numerical means are required. For AdS-Milne, it is particularly easy to solve the equations, calculate
the regulated length and obtain the correlator. The result matches with (\ref{onetwo}) in the large $\Delta$ limit 
for the space-like separated points. We do not reproduce the calculation here. 

As a nontrivial example, let us next look at the case of $p_1 = p_2 = 1/2$ and $p_3 = 0$.
Metric now has a true singularity in the past. Here (\ref{complitwo}) and (\ref{complifive}) can be explicitly 
evaluated. The solutions are
\begin{eqnarray}
&&\bar x^1 (\bar t) = \pm {\sqrt{2}} \Big(~\rm{sin}^{-1} {\sqrt {\bar t}} - \frac{\pi}{2}\Big),\nonumber\\
&&\bar x^2 (\bar t) = \pm {\sqrt{2}} \Big(~\rm{sin}^{-1} {\sqrt {\bar t}} - \frac{\pi}{2}\Big),\nonumber\\
&&\bar x^3 (\bar t) = \pm \frac{1}{\sqrt 2}(\frac{\pi}{2} - {\sqrt{\bar t - \bar t^2}}  ~\rm{sin}^{-1}{\sqrt {\bar t}}),
\label{xsol}
\end{eqnarray}
and
\begin{eqnarray}
&&\bar z(\bar t) =  \frac{1}{\sqrt 2}\Big[ (\bar t - \bar t_0)(\bar t + \bar t_0 -1) - 2 {\sqrt{(1 - \bar t) \bar t}}
~{\rm{sin}}^{-1} {\sqrt{ 1- \bar t}} - ~({\rm{sin}}^{-1} {\sqrt{ 1- \bar t}})^2 \nonumber\\
&&~~~~~~~~~~~~~~+  ~({{\rm{sin}}^{-1}} {\sqrt{ 1- \bar t_0}})^2
+ 2 {\sqrt{(1 - \bar t_0) \bar t_0}}~{\rm{sin}}^{-1} {\sqrt{ 1- \bar t_0}}\Big]^{\frac{1}{2}}.
\end{eqnarray}
Here we have used the scaling symmetry of (\ref{scal}) with $\lambda = t^*$ to set the turning point to one. The
parameter $t_0$ appearing in the above equation is the space-like surface on which the correlator would
be calculated.

\bigskip

\noindent{\underbar{\bf Correlators within the geodesic approximation}}

\bigskip
 
These solutions can now be used to compute the regularized geodesic length by subtracting the equivalent AdS part
from
\begin{equation}
{\cal L} = - \int_{\bar t = 1}^{\bar t= \bar t_0 + \bar\delta} \frac{2 d\bar t}{\bar z}\Big[{\sqrt{
-1 + \Big(\frac{d\bar z}{d\bar t}\Big)^2 + \sum_{j = 1}^3 \bar{t}^{2 p_j} \Big(\frac{d\bar x^j}{d \bar t}\Big)^2    }}\Big]
\label{scalone}
\end{equation}
and then to use it in (\ref{cor}) to evaluate the correlator.   
It turns out that, for the situation we are considering, the integral can not be done analytically. 
However, we obtain it numerically and the result is shown in 
the figure. What we see is that the correlator goes to zero as we take the space-like surface $t_0$ close to
the initial singularity.
\begin{figure}[H]
\label{fig1}
 \centering
 \includegraphics[width=.5\textwidth]{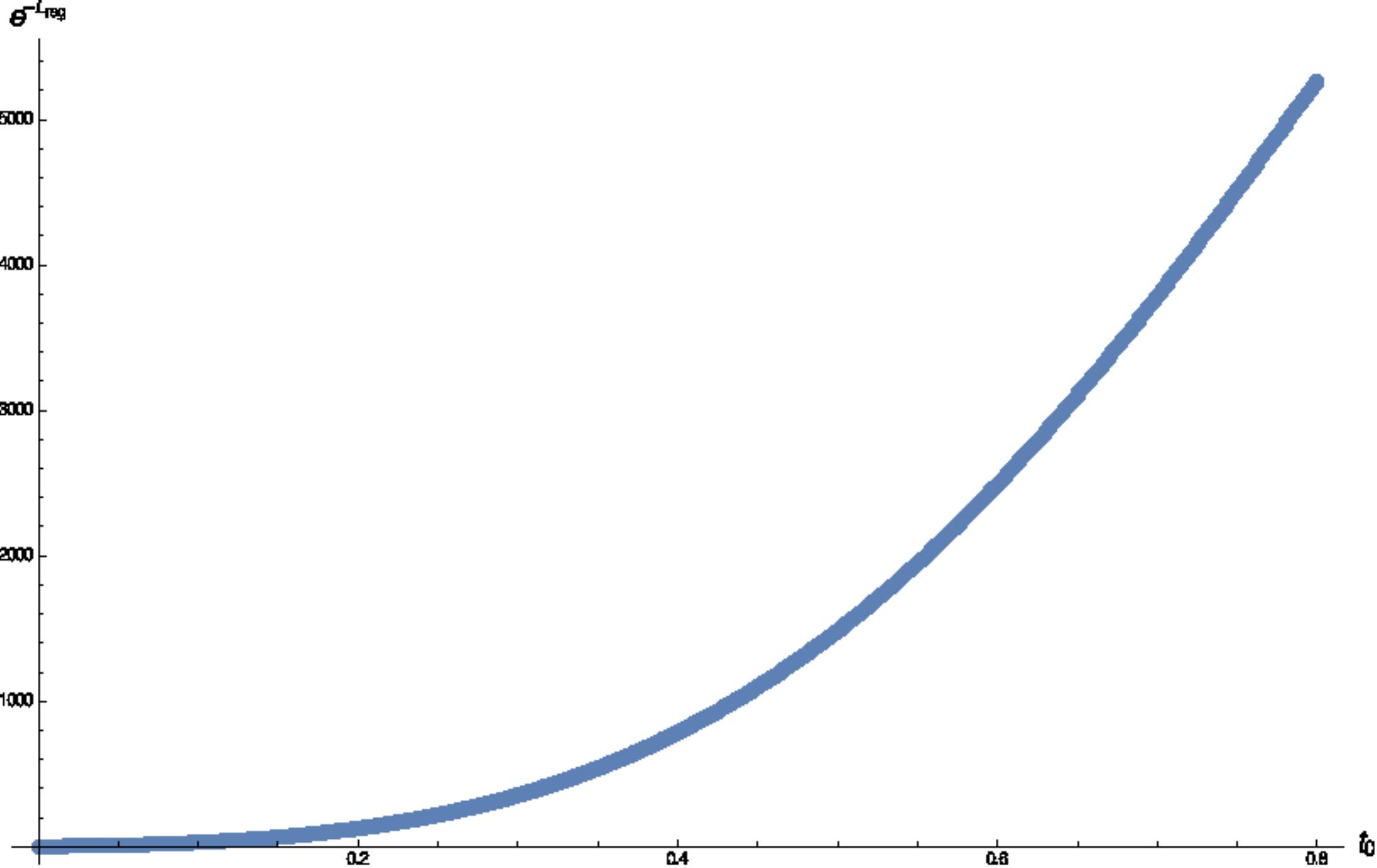}
 \caption{Plot of $e^{-L_{reg}}$ as a function of $t_0$ for $p_1 = p_2 = \frac{1}{2}, p_3 = 0$ non-vacuum AdS-Kasner solution. 
As the space-like surface at $t_0$ is pushed towards $t = 0$, the correlator vanishes.}
\end{figure}  

For the Kasner-AdS soliton (\ref{ksoliton}), where the $z \rightarrow \infty$ singularity is removed by adding 
a periodic coordinate, one can also compute the correlator in the geodesic approximation. The qualitative results remain unchanged. 
For the illustrative purpose, we do it here for $p_1 = p_2 = 1/2, p_3 = 0$. While the solutions for the geodesic 
equations for $x^i$s remain same as  (\ref{xsol}), the one for the $z$ changes. It is given by
\begin{equation}
\ddot z + \Gamma_{tt}^{z} + \sum_{i=1}^3 \Gamma_{x_i x_i}^z \dot x_i^2 + (\Gamma_{zz}^z + \frac{2}{z})\dot z^2 - 
\frac{1}{2}\dot x_1^2 \dot z  - \frac{1}{2}\dot x_2^2 \dot z = 0,
\label{zeq}
\end{equation}
where
\begin{eqnarray}
&&\Gamma_{tt}^{z} =  -\Gamma_{x_3 x_3}^z = \frac{(z - z_0) (z^4 + z^3 z_0 + z^2 z_0^2 + z z_0^3 + z_0^4)}{z z_0^5},\nonumber\\
&& \Gamma_{x_1 x_1}^z =  \Gamma_{x_2 x_2}^z = t \Gamma_{x_3 x_3}^z,\nonumber\\
&&\Gamma_{zz}^z = - \frac{7 z^5 - 2 z_0^5}{2 z (z - z_0) (z^4 + z^3 z_0 + z^2 z_0^2 + z z_0^3 + z_0^4)}.
\end{eqnarray}
In the above computations we have turned off the periodic coordinate $\theta$ in the metric (\ref{ksoliton})
The length of the geodesic now reads
\begin{eqnarray}
&&{\cal L} = -2 \int \frac{dx}{z(x)} \frac{{\rm sin}({\sqrt 2} x)}{\sqrt 2}\nonumber\\ 
&&~~~~~~~~~~~~~~~\Big[ \frac{2}{f(z) ~{\rm sin}^2 ( {\sqrt 2} x)} z^\prime (x)^2 - 1
+ 4 ~{\rm cos}^2 (\frac{x}{\sqrt 2}) ~{\rm cosec}^2 ( {\sqrt 2} x) + ~{\rm cot}^2 (\frac{x}{\sqrt 2})\Big]^{\frac{1}{2}},
\label{soll}
\end{eqnarray}
where $f(z) = 1 - z^5/z_0^5$ and  the prime denotes derivative with respect to the argument. Equation (\ref{zeq}) and the 
integral in (\ref{soll}) can be solved numerically.The result of the correlator as a function of $t_0$ is shown in figure 2. 
\begin{figure}[H]
\label{fig2}   
 \centering
 \includegraphics[width=.50\textwidth]{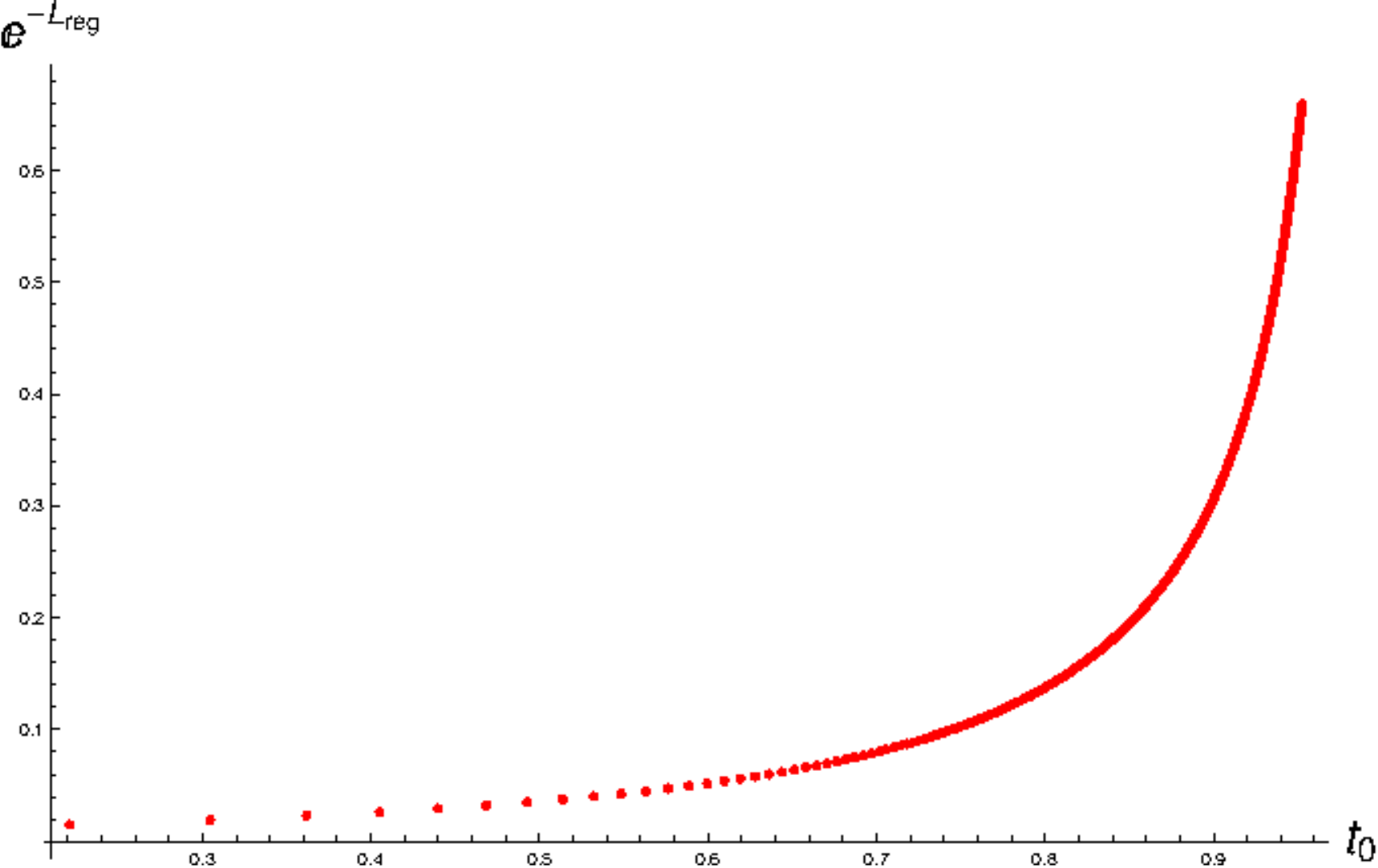}
 \caption{Plot of $e^{-L_{reg}}$ as a function $t_0$ for the soliton solution.}
\end{figure}

While we do not have a general proof, we believe that, within the approximations used here, for general non-negative $p_i$s,
the spacelike two point correlator is nonsingular even if we push the surface close to its past singularity. It is 
perhaps prudent at this point  to compare our results with the existing ones. In \cite{Engelhardt:2014mea}, \cite{Engelhardt:2015gta}, the correlators
are computed for space-like separated operators along the direction with $p_i < 0$. It is found that the correlators
become singular as the geodesics approach the past singularity. This blowing up of the space-like correlator is subsequently
argued to be a consequence of boundary conformal field theory picking up a non-normalizable state. In the present context, 
for $p_i > 0$, we see that the geodesic approximation selects a perfectly normalizable CFT state with the usual behaviour
of space-like correlator.

We end this section by pointing out a Ward identity\footnote{ This identity is related to the 
existence of conformal Killing vector  $K = t\partial_t +\sum_{i=1}^3 (1 - p_i) x^i \partial_{i} $ for the Kasner 
metric}satisfied by the correlator. It arises very simply from  
our previous analysis. For general $p_i$s, the expression which determines the correlator is the 
integral given in (\ref{scalone}). Due to the bulk symmetry (\ref{scal}), this expression,
once evaluated, depends only on the scaled time $\bar t_0$. Therefore, we write
\begin{equation}
\langle {\cal O}(x_1, x_2, x_3, t) {\cal O}(-x_1, -x_2, -x_3, t) \rangle = f(\bar{t}_0),
\end{equation}
where $f(\bar{t}_0)$ is purely a function of $\bar{t}_0$. In other words, 
\begin{equation}
\langle {\cal O}(\lambda^{1-p_i}\bar x_i, \lambda \bar t) {\cal O}(-\lambda^{1-p_i} \bar x_i, \lambda \bar t) \rangle = 
f(\bar{t}_0).
\end{equation}
Now differentiating both sides with respect to $\lambda$ and setting $\lambda = 1$, we immediately get
\begin{equation}
\Big(\sum_{i=1}^3 (1 - p_i) \frac{\partial}{\partial ~{\rm log}~\bar x_i} +  \frac{\partial}{\partial ~{\rm log}~\bar t}\Big)
\langle {\cal O}(\bar x_i, \bar t) {\cal O}(-\bar x_i, \bar t) \rangle  = 0.
\end{equation}

\section{Discussion}

In this work, we have carried out several investigations on AdS cosmologies and their gauge theory duals. We analyzed 
various time dependent geometries in AdS, in particular non-vacuum AdS-Kasner  and AdS-Kasner soliton and studied 
general spacelike geodesics and scalar 
fields in these. Using gauge-gravity duality, we calculated two point correlators and studied their behaviour near the 
singularity. Even though the bulk cantains space-like sigularity which extends all the way to the boundary, 
for positive Kasner exponents, $p_i > 0$, we found that, within the geodesic approximation,  
the space-like seperated correlators vanish as the geodesic approaches the singularity. This perhaps suggests
that, regardless of the presence of the past singulatity, the boundary CFT state 
may be a perfectly normalizable one. This is unlike the cases studied in  \cite{Engelhardt:2014mea}, \cite{Engelhardt:2015gta},
where for $p_i < 0$, the geodesic approximation seems to pick a non-normalizabe CFT state causing a divergence
in space-like correlator. In this paper, for characterizing these geometries, we also embedded these metrics in higher dimensional 
flat space and found corresponding equations for surfaces. These may turn out to be useful in future studies. 

There are several avenues for future explorations. It might be worth studying features of strongly coupled gauge theories in 
other time dependent backgrounds.
Indeed it is possible to construct a large class of time dependent backgrounds with bulk duals. Consider, for 
example, the Taub metric belonging to the Bianchi type II spacetime \cite{wainwright}: 
\begin{equation}
dS^2 = - A(t)^2 dt^2 + t^{2p_1} A(t)^{-2} ( dx_1 + 4 p_1 b x_3 dx_2)^2 + t^{2p_2} A(t)^2 dx_2^2 + t^{2p_3} A(t)^2 dx_3^2,
\end{equation}
with
\begin{equation}
A(t) = {\sqrt{1 + b^2 t^{4p_1}}}, ~~\sum_{i=1}^3 p_i = 1, ~~\sum_{i=1}^3 {p_i}^2 = 1,
\end{equation}
where $b$ is a constant. The metric is Ricci-flat and can therefore be embedded in AdS. The 
anisotropic scaling symmetry of AdS-Kasner is broken by the parameter $b$. 
One may wish to compute the boundary correlator of higher dimensional
operators which are spacelike separated along $x_1$ direction. In this regard, the case of $p_1 = 1, p_2 =0, p_3 = 0$ is 
particularly interesting and simple. For $b = 0$, it is a Milne metric. For non-zero
$b$, it is a deformation of Milne. While at $t = 0$, the  Kretschmann scalar reaches a $b$ dependent
constant, it is zero for large $t$. Computation of the correlator in the geodesic approximation 
should be particularly easy in this case. Since for the Milne, the field theory state is a thermal one with temperature 
$1/(2\pi)$, one may wish to know as to what happens if $b$ is tuned. 

Another example which one many wish to explore is AdS-Kantowski-Sachs metric. One example of vacuum Kantowski-Sachs metric 
is the interior Schwarzschild metric. Many of our calculations should be easily extendable to  this case also.

\bigskip

{\bf Acknowledgment:} We thank Shamik Banerjee, Souvik Banerjee, Samrat Bhowmick and Amitabh Virmani for 
useful discussions. The works of SC and SPC are supported in part by the DST-Max Planck Partner
Group ``Quantum Black Holes'' between IOP Bhubaneswar and AEI Golm.


\begin{thebibliography}{99}

%\cite{Kraus:2002iv}
\bibitem{Kraus:2002iv} 
  P.~Kraus, H.~Ooguri and S.~Shenker,
  ``Inside the horizon with AdS / CFT,''
  Phys.\ Rev.\ D {\bf 67}, 124022 (2003)
  doi:10.1103/PhysRevD.67.124022
  [hep-th/0212277].
  %%CITATION = doi:10.1103/PhysRevD.67.124022;%%
  
  %\cite{Koyama:2001rf}
\bibitem{Koyama:2001rf}
  K.~Koyama and J.~Soda,
  ``Strongly coupled CFT in FRW universe from AdS / CFT correspondence,''
  JHEP {\bf 0105}, 027 (2001)
  [hep-th/0101164].
  %%CITATION = HEP-TH/0101164;%%
  %31 citations counted in INSPIRE as of 14 Apr 2015

%\cite{Hertog:2004rz}
\bibitem{Hertog:2004rz} 
  T.~Hertog and G.~T.~Horowitz,
  ``Towards a big crunch dual,''
  JHEP {\bf 0407}, 073 (2004)
  [hep-th/0406134].
  %%CITATION = HEP-TH/0406134;%%
  %139 citations counted in INSPIRE as of 29 Oct 2015



%\cite{Hertog:2005hu}
\bibitem{Hertog:2005hu} 
  T.~Hertog and G.~T.~Horowitz,
  ``Holographic description of AdS cosmologies,''
  JHEP {\bf 0504}, 005 (2005)
  [hep-th/0503071].
  %%CITATION = HEP-TH/0503071;%%

%\cite{Das:2006dz}
\bibitem{Das:2006dz} 
  S.~R.~Das, J.~Michelson, K.~Narayan and S.~P.~Trivedi,
  %``Time dependent cosmologies and their duals,''
  Phys.\ Rev.\ D {\bf 74}, 026002 (2006)
  [hep-th/0602107].
  %%CITATION = HEP-TH/0602107;%%
  %76 citations counted in INSPIRE as of 29 Oct 2015


%\cite{Awad:2007fj}
\bibitem{Awad:2007fj}
  A.~Awad, S.~R.~Das, K.~Narayan and S.~P.~Trivedi,
  ``Gauge theory duals of cosmological backgrounds and their energy momentum tensors,''
  Phys.\ Rev.\ D {\bf 77}, 046008 (2008)
  [arXiv:0711.2994 [hep-th]].
  %%CITATION = ARXIV:0711.2994;%%
  %26 citations counted in INSPIRE as of 25 Nov 2014
  

%\cite{Awad:2008jf}
\bibitem{Awad:2008jf} 
  A.~Awad, S.~R.~Das, S.~Nampuri, K.~Narayan and S.~P.~Trivedi,
  ``Gauge Theories with Time Dependent Couplings and their Cosmological Duals,''
  Phys.\ Rev.\ D {\bf 79}, 046004 (2009)
  [arXiv:0807.1517 [hep-th]].
  %%CITATION = ARXIV:0807.1517;%%


%\cite{Awad:2009bh}
\bibitem{Awad:2009bh} 
  A.~Awad, S.~R.~Das, A.~Ghosh, J.~-H.~Oh and S.~P.~Trivedi,
  ``Slowly Varying Dilaton Cosmologies and their Field Theory Duals,''
  Phys.\ Rev.\ D {\bf 80}, 126011 (2009)
  [arXiv:0906.3275 [hep-th]].
  %%CITATION = ARXIV:0906.3275;%%

%\cite{Craps:2007ch}
\bibitem{Craps:2007ch}
  B.~Craps, T.~Hertog and N.~Turok,
  ``On the Quantum Resolution of Cosmological Singularities using AdS/CFT,''
  Phys.\ Rev.\ D {\bf 86}, 043513 (2012)
  [arXiv:0712.4180 [hep-th]].
  %%CITATION = ARXIV:0712.4180;%%
  
%\cite{Banerjee:2013jn}
\bibitem{Banerjee:2013jn} 
  S.~Banerjee, S.~Bhowmick and S.~Mukherji,
  ``Anisotropic branes,''
  Phys.\ Lett.\ B {\bf 726}, 461 (2013)
  [arXiv:1301.7194 [hep-th]].
  %%CITATION = ARXIV:1301.7194;%%
  %1 citations counted in INSPIRE as of 29 Oct 2015


\bibitem{Pedraza:2013nlk} 
  W.~Fischler, S~Kundu, J.~F.~Pedraza,
  ``Entanglement and out-of-equilibrium dynamics in holographic models of de Sitter QFTs'',
    JHEP {\bf 07}, 021 (2014) 
   [arXiv:1311.5519 [hep-th]].    
  
  
\bibitem{Engelhardt:2014mea} 
  N.~Engelhardt, T.~Hertog and G.~T.~Horowitz,
  ``Holographic Signatures of Cosmological Singularities,''
  Phys.\ Rev.\ Lett.\  {\bf 113}, 121602 (2014)
  [arXiv:1404.2309 [hep-th]].
  %%CITATION = ARXIV:1404.2309;%%
  %1 citations counted in INSPIRE as of 28 Oct 2014

\bibitem{Pedraza:2014ytc} 
  J.~F.~Pedraza,
  ``Evolution of non-local observables in an expanding boost-invariant plasma,''
   [arXiv:1405.1724 [hep-th]].  
  
%\cite{Engelhardt:2015gta}
\bibitem{Engelhardt:2015gta} 
  N.~Engelhardt, T.~Hertog and G.~T.~Horowitz,
  ``Further Holographic Investigations of Big Bang Singularities,''
  JHEP {\bf 1507}, 044 (2015)
  [arXiv:1503.08838 [hep-th]].
  %%CITATION = ARXIV:1503.08838;%%

%\cite{Banerjee:2015fua}
\bibitem{Banerjee:2015fua} 
  S.~Banerjee, S.~Bhowmick, S.~Chatterjee and S.~Mukherji,
  ``A note on AdS cosmology and gauge theory correlator,''
  JHEP {\bf 1506}, 043 (2015)
  [arXiv:1501.06317 [hep-th]].
  %%CITATION = ARXIV:1501.06317;%%
  %1 citations counted in INSPIRE as of 20 août 2015

%\cite{Engelhardt:2015gla}
\bibitem{Engelhardt:2015gla} 
  N.~Engelhardt and G.~T.~Horowitz,
  ``Holographic Consequences of a No Transmission Principle,''
  arXiv:1509.07509 [hep-th].
  %%CITATION = ARXIV:1509.07509;%%

%\cite{Kumar:2015jxa}
\bibitem{Kumar:2015jxa} 
  S.~P.~Kumar and V.~Vaganov,
  ``Probing crunching AdS cosmologies,''
  arXiv:1510.03281 [hep-th].
  %%CITATION = ARXIV:1510.03281;%%

%\cite{Brandenberger:2016egn}
\bibitem{Brandenberger:2016egn} 
  R.~H.~Brandenberger, Y.~F.~Cai, S.~R.~Das, E.~G.~M.~Ferreira, I.~A.~Morrison and Y.~Wang,
  %``Fluctuations in a Cosmology with a Space-Like Singularity and their Gauge Theory Dual Description,''
  arXiv:1601.00231 [hep-th].
  %%CITATION = ARXIV:1601.00231;%%
  %5 citations counted in INSPIRE as of 27 Aug 2016

%\cite{Deger:2016axo}
\bibitem{Deger:2016axo} 
  N.~S.~Deger,
  ``Time-Dependent AdS Backgrounds from S-Branes,''
  doi:10.1016/j.physletb.2016.07.024
  arXiv:1606.00674 [hep-th].
  %%CITATION = doi:10.1016/j.physletb.2016.07.024;%%

%\cite{Horowitz:1998ha}
\bibitem{Horowitz:1998ha} 
  G.~T.~Horowitz and R.~C.~Myers,
  %``The AdS / CFT correspondence and a new positive energy conjecture for general relativity,''
  Phys.\ Rev.\ D {\bf 59}, 026005 (1998)
  doi:10.1103/PhysRevD.59.026005
  [hep-th/9808079].
  %%CITATION = doi:10.1103/PhysRevD.59.026005;%%
  %307 citations counted in INSPIRE as of 07 Jan 2017

\bibitem{Haehl:2012fh}  
    Felix~ M.~ Haehl,
   ``The Schwarzschild-Black String AdS Soliton: Instability and Holographic Heat Transport, ''
   Class.\ Quant. \ Grav. {\bf 30}, 055002 (2013),
   arXiv:1210.5763[hep-th].

  
%\cite{Erickson:2003zm}
\bibitem{Erickson:2003zm} 
  J.~K.~Erickson, D.~H.~Wesley, P.~J.~Steinhardt and N.~Turok,
  ``Kasner and mixmaster behavior in universes with equation of state $w \ge 1$,''
  Phys.\ Rev.\ D {\bf 69}, 063514 (2004)
  [hep-th/0312009].
  %%CITATION = HEP-TH/0312009;%%
  %143 citations counted in INSPIRE as of 20 Aug 2015

%\cite{Fischetti:2014hxa}
\bibitem{Fischetti:2014hxa} 
  S.~Fischetti, D.~Kastor and J.~Traschen,
  ``Non-Vacuum AdS Cosmologies and the Approach to Equilibrium of Entanglement Entropy,''
  Class.\ Quant.\ Grav.\  {\bf 31}, no. 23, 235007 (2014)
  [arXiv:1407.4299 [hep-th]].
  %%CITATION = ARXIV:1407.4299;%%
  %2 citations counted in INSPIRE as of 20 août 2015

%\cite{Deser:1997ri}
\bibitem{Deser:1997ri} 
  S.~Deser and O.~Levin,
  %``Accelerated detectors and temperature in (anti)-de Sitter spaces,''
  Class.\ Quant.\ Grav.\  {\bf 14}, L163 (1997)
  doi:10.1088/0264-9381/14/9/003
  [gr-qc/9706018].
  %%CITATION = doi:10.1088/0264-9381/14/9/003;%%
  %131 citations counted in INSPIRE as of 07 Jan 2017

%\cite{Deser:1998bb}
\bibitem{Deser:1998bb} 
  S.~Deser and O.~Levin,
  %``Equivalence of Hawking and Unruh temperatures through flat space embeddings,''
  Class.\ Quant.\ Grav.\  {\bf 15}, L85 (1998)
  doi:10.1088/0264-9381/15/12/002
  [hep-th/9806223].
  %%CITATION = doi:10.1088/0264-9381/15/12/002;%%
  %64 citations counted in INSPIRE as of 07 Jan 2017

%\cite{Deser:1998xb}
\bibitem{Deser:1998xb} 
  S.~Deser and O.~Levin,
  %``Mapping Hawking into Unruh thermal properties,''
  Phys.\ Rev.\ D {\bf 59}, 064004 (1999)
  doi:10.1103/PhysRevD.59.064004
  [hep-th/9809159].
  %%CITATION = doi:10.1103/PhysRevD.59.064004;%%
  %115 citations counted in INSPIRE as of 07 Jan 2017

%\cite{Kehagias:2013xga}
\bibitem{Kehagias:2013xga} 
  A.~Kehagias and A.~Riotto,
  %``Conformal Symmetries of FRW Accelerating Cosmologies,''
  Nucl.\ Phys.\ B {\bf 884}, 547 (2014)
  doi:10.1016/j.nuclphysb.2014.05.006
  [arXiv:1309.3671 [hep-th]].
  %%CITATION = doi:10.1016/j.nuclphysb.2014.05.006;%%
  %10 citations counted in INSPIRE as of 07 Jan 2017



  
\bibitem{blau:lec}
Matthias~Blau,``Lecture Notes on General Relativity'',\\
http://www.blau.itp.unibe.ch/Lecturenotes.html  
  
\bibitem{wainwright}
   J. Wainwright and G.F.R. Ellis, ``Dynamical systems in Cosmology,'' Cambridge University Press, 
    1997, page 196.

%\cite{Paston:2013uia}
\bibitem{Paston:2013uia} 
  S.~A.~Paston and A.~A.~Sheykin,
  %``Embeddings for solutions of Einstein equations,''
  Theor.\ Math.\ Phys.\  {\bf 175}, 806 (2013)
  doi:10.1007/s11232-013-0067-4
  [arXiv:1306.4826 [gr-qc]].
  %%CITATION = doi:10.1007/s11232-013-0067-4;%%
  %4 citations counted in INSPIRE as of 27 Aug 2016

%\cite{Louko:2000tp}
\bibitem{Louko:2000tp} 
  J.~Louko, D.~Marolf and S.~F.~Ross,
  %``On geodesic propagators and black hole holography,''
  Phys.\ Rev.\ D {\bf 62}, 044041 (2000)
  doi:10.1103/PhysRevD.62.044041
  [hep-th/0002111].
  %%CITATION = doi:10.1103/PhysRevD.62.044041;%%
  %95 citations counted in INSPIRE as of 27 Aug 2016

%\cite{Fidkowski:2003nf}
\bibitem{Fidkowski:2003nf} 
  L.~Fidkowski, V.~Hubeny, M.~Kleban and S.~Shenker,
  %``The Black hole singularity in AdS / CFT,''
  JHEP {\bf 0402}, 014 (2004)
  doi:10.1088/1126-6708/2004/02/014
  [hep-th/0306170].
  %%CITATION = doi:10.1088/1126-6708/2004/02/014;%%
  %213 citations counted in INSPIRE as of 27 Aug 2016

%\cite{Festuccia:2005pi}
\bibitem{Festuccia:2005pi} 
  G.~Festuccia and H.~Liu,
  %``Excursions beyond the horizon: Black hole singularities in Yang-Mills theories. I.,''
  JHEP {\bf 0604}, 044 (2006)
  doi:10.1088/1126-6708/2006/04/044
  [hep-th/0506202].
  %%CITATION = doi:10.1088/1126-6708/2006/04/044;%%
  %92 citations counted in INSPIRE as of 27 Aug 2016



%############################################
\end{thebibliography}
\end{document}